\documentclass[lettersize,journal]{IEEEtran}
\usepackage{amsmath,amsfonts}
\usepackage{ragged2e}
\usepackage{siunitx}
\usepackage{caption}
\usepackage{amsmath}
\usepackage{amssymb}
\usepackage{array}
\usepackage{textcomp}
\usepackage{stfloats}
\usepackage{url}
\usepackage{adjustbox}
\usepackage{verbatim}
\usepackage{graphicx}
\usepackage{cite}
\usepackage{tikz}
\usepackage[flushleft]{threeparttable}
\usepackage{amsthm}
\usepackage{subcaption}
\usepackage[normalem]{ulem}
\usepackage{etoolbox}
\usepackage{amssymb}
\newtheorem{remark}{Remark}

\usepackage{mathtools}

\usepackage{breqn}
\usepackage[multiple]{footmisc}

\definecolor{ForestGreen}{RGB}{34,139,34}
\usepackage{bbm}

\DeclareMathOperator*{\argmax}{arg\,max}

\allowdisplaybreaks

\usepackage{algorithm}
\usepackage{algpseudocode}
\usepackage{setspace}
\usepackage[font=footnotesize,labelfont=bf,skip=5pt]{caption}
\usepackage{subcaption}
\usepackage{enumitem}
\usepackage{bigints}
\usepackage[colorlinks]{hyperref}
\usepackage{flushend}
\hypersetup{
	urlcolor     = magenta, 
	linkcolor    = blue, 
	filecolor	    = cyan, 
	citecolor   = {blue}, 
}

\usepackage{scalerel}[2016-12-29]

\usepackage{wrapfig}
\usepackage[outdir=./]{epstopdf}
\usepackage{multirow}
\begin{document}
\title{\vspace{-0.1cm} Practical RIS Gain without the Pain: Randomization and Opportunistic Scheduling in 5G NR}
\author{L. Yashvanth,~\emph{Student Member, IEEE}, Raju Malleboina,~\emph{Student Member, IEEE}, Venkatareddy Akumalla,~\emph{Member, IEEE}, Nekkanti Guna Sai Kiran, Debdeep Sarkar,~\emph{Senior Member, IEEE}, and Chandra R. Murthy,~\emph{Fellow, IEEE}
\thanks{The authors are with the Dept. of ECE, Indian Institute of Science (IISc), Bengaluru, India 560012. E-mails: \{yashvanthl, malleboinar\}@iisc.ac.in, venkatareddy@fsid-iisc.in, \{gunanekkanti, debdeep, cmurthy\}@iisc.ac.in.} 
\thanks{\textcolor{black}{This work was financially supported in parts by a Qualcomm $6$G UR grant, and the Indian Space Research Organization (ISRO), Govt. of India, and Defense Research and Development Organization (DRDO) Grants numbered ISTC/EEC/DS/492 and ADE/CARS/24-25/02, respectively. The authors thank Prof. K. J. Vinoy, Dept. of ECE, IISc, Bengaluru for providing directional horn antennas for the experiments.}}
\vspace{-0.4cm}
}
 \maketitle
\begin{abstract}
We experimentally demonstrate the performance gains achieved by an in-house built reconfigurable intelligent surface (RIS) integrated with a real-time 5G new radio (NR) system implemented using the OpenAirInterface (OAI) framework. We first quantify the gain in throughput achievable by integrating an RIS with a 5G system. Next, we show that randomly setting the RIS phase configuration and leveraging the inherent proportional fair (PF) scheduling mechanism of 5G NR can yield near-optimal throughput, provided the throughput averaging window of the PF scheduler is chosen judiciously. 
This occurs because, in each time slot, the PF scheduler naturally prioritizes data transmission to the user equipment (UE) that experiences the best channel conditions, namely, the UE to which the randomly configured RIS is aligned. Subsequently, we experimentally evaluate key performance metrics, including the reference signal received power (RSRP), block error rate (BLER), modulation and coding scheme (MCS) index, and throughput, under random RIS configurations. These results confirm that even a randomly configured RIS with negligible overhead can deliver performance comparable to optimized RIS designs, in real-world 5G NR wireless communication systems.

\end{abstract}
\begin{IEEEkeywords}
Reconfigurable intelligent surfaces, 5G NR, Proportional-fair scheduling, Experimental validation.
\end{IEEEkeywords}

\color{black}
\vspace{-0.2cm}
\section{Introduction}
Reconfigurable intelligent surface (RIS) has emerged as a promising means to enhance next-generation wireless networks, by allowing one to control the radio propagation environment~\cite{Tang_JSAC_2020,Linglong_Exp_Access_2020}. By appropriately configuring the phase shifts of the RIS elements, significant throughput gains can be achieved. However, these gains come at the cost of high overhead due to channel state information (CSI) estimation, phase optimization, and control signaling steps. Moreover, the practical integration of RIS hardware with real-time $5$G new radio (NR) systems, and the experimental validation of the resulting gains has received very limited attention in the literature. In this paper, we experimentally demonstrate the benefits of integrating an RIS with a real-time $5$G NR testbed. We show that \emph{a randomly configured RIS  combined with the inherent opportunistic scheduling scheme of $5$G NR achieves near-optimal performance, thereby avoiding the associated optimization and signaling overheads.}

While extensive theoretical research has been conducted on RIS, only a few experimental studies exist. In particular,~\cite{Tang_JSAC_2020} proposed amplitude-and-phase-varying modulation for RISs;~\cite{Linglong_Exp_Access_2020} developed a $2$-bit RIS prototype;~\cite{Lin_Exp_TCOM_2024} built and optimized a realistic reflection model for RIS;~\cite{Yanqing_Exp_WCL_2023} investigated RIS deployment strategies;~\cite{Emil_Exp_TCOM_2021} demonstrated indoor and outdoor gains using an optimized RIS;~\cite{Wankai_Exp_TWC_2021} modeled path loss via an RIS;~\cite{lodro2022experimental} analyzed the impact of RIS on performance of mobile operators;~\cite{Kayraklik_Exp_ICC_2023} demonstrated coverage extension due to an RIS; and~\cite{Linglong_Exp_TCOM_2023} compared active and passive RIS elements. However, these studies employ vector signal generators, vector network analyzers (VNA), or software-defined radios (SDR), and do not assess performance in a fully integrated real-time $5$G NR testbed. A setup involving a $5$G testbed would closely resemble the gains obtained in practical deployment, where interactions between RIS control and $5$G protocols involving base station (BS)–user equipment (UE) connection establishment, selection of modulation and coding scheme (MCS), resource allocation, CSI estimation overheads, etc., are accounted for. In this regard,~\cite{sahin2025ris} examines an RIS in a real-time $5$G setup, but under CSI-dependent optimized phase configuration. However, as indicated before, optimization-based methods incur large implementation overheads, reducing the benefits from RIS.

In this work, we propose and experimentally evaluate a low-complexity approach that bypasses optimization of RIS phases. In our setup, the RIS phase coefficients randomly switch between different states, while the built-in opportunistic scheduling algorithms at the $5$G medium access control (MAC) layer select the UE experiencing the best instantaneous channel. This combination effectively exploits multi-user diversity, allowing the system to capture most of the RIS beamforming benefits without explicit per-element CSI estimation/RIS optimization. The key contributions of our paper are as follows:
\begin{enumerate}[leftmargin=*]
\item We first develop a real-time $5$G NR testbed using an OpenAirInterface (OAI) framework that is integrated with an in-house-built RIS and demonstrate that the presence of an RIS yields better performance than its absence.
\item We investigate the temporal variation of key performance metrics, including reference signal received power (RSRP), MCS index, block error rate (BLER), and throughput at different UEs under randomly configured RIS, and when the $5$G BS uses a proportional fair (PF) scheduler.
\item Finally, we experimentally show that a randomly configured RIS can achieve performance close to that of optimized RIS schemes by leveraging the inherent opportunistic schedulers of $5$G NR provided the averaging window size of the PF scheduler is chosen appropriately.
\end{enumerate}
Thus, our experiments provide a practical proof of concept that integrating an RIS need not add substantial complexity to a $5$G system. Randomized RIS configurations, when coupled with the native scheduling mechanisms of $5$G NR, is sufficient to deliver the benefits attributed to optimized RIS designs.

\section{Description of the RIS-Aided $5$G NR System}

The BS (called a gNB in $5$G NR) serves $K$ UEs, aided by an $N$-element RIS. Let $\mathbf{h}_1 \in \mathbb{C}^{N}$ and $\mathbf{h}_{2,k} \in \mathbb{C}^{N}$ denote the BS-RIS and RIS-UE-$k$ channel, respectively. Then, the received signal at UE-$k$ in time $t$ is given by\footnote{The direct BS-UE link is neglected for simplicity.}
\begin{align}
    y_k(t) &= \sum\nolimits_{n=1}^{N}\left[\mathbf{h}_{2,k}\right]_n\left[\mathbf{h}_{1}\right]_n e^{j\theta_n(t)} x_k(t) + n_k(t)\\
    &= \mathbf{h}_{2,k}^T\boldsymbol{\Phi}(t)\mathbf{h}_1x_k(t) + n_k(t), \label{eq_channel_basic}
\end{align}
where $\boldsymbol{\Phi}(t)$ is a diagonal matrix of RIS phase shifts: $\{e^{j\theta_n}\}_{n=1}^{N}$, $x_k(t)$ is the transmitted symbol, and $n_k(t)$ is the additive noise at UE-$k$, at time $t$.
Let the cascaded BS-RIS-UE channel be $\mathbf{h}_{c,k} \triangleq \mathbf{h}_1 \odot \mathbf{h}_{2,k}$ and the (conjugated) RIS phase vector be $\boldsymbol{\phi}(t) \triangleq [e^{-j\theta_1(t)},\ldots,e^{-j\theta_N(t)}]^T$, where $\odot$ is the Hadamard product. Then, the received signal in~\eqref{eq_channel_basic} becomes
\vspace{-0.1cm}
\begin{equation}
\vspace{-0.1cm}
{y}_k(t) = \boldsymbol{\phi}^H(t)\mathbf{h}_{c,k} x_k(t) + n_k(t),
\end{equation}
and we can express the effective channel to UE-$k$ at time $t$ as
\vspace{-0.2cm}
\begin{equation}\label{eq_overall_ch_UE_k}
\vspace{-0.1cm}
    h_k(t) = \boldsymbol{\phi}^H(t)\mathbf{h}_{c,k}.
\end{equation}
Next, we describe the hardware and software setup used to demonstrate the real-time gains of a randomly configured RIS. It involves an \textcolor{black}{RIS} prototype (Sec.~\ref{sec_RIS_design}) and an OAI-based implementation of a 3GPP-compliant $5$G NR radio access network (RAN) protocol stack (Sec.~\ref{sec_OAI_description}).
\vspace{-0.25cm}
\subsection{Design and Modeling of the In-house-Built RIS}\label{sec_RIS_design}

\begin{figure}[t]
\vspace{-0.2cm}
        \centering
\includegraphics[width=0.9\columnwidth,keepaspectratio]{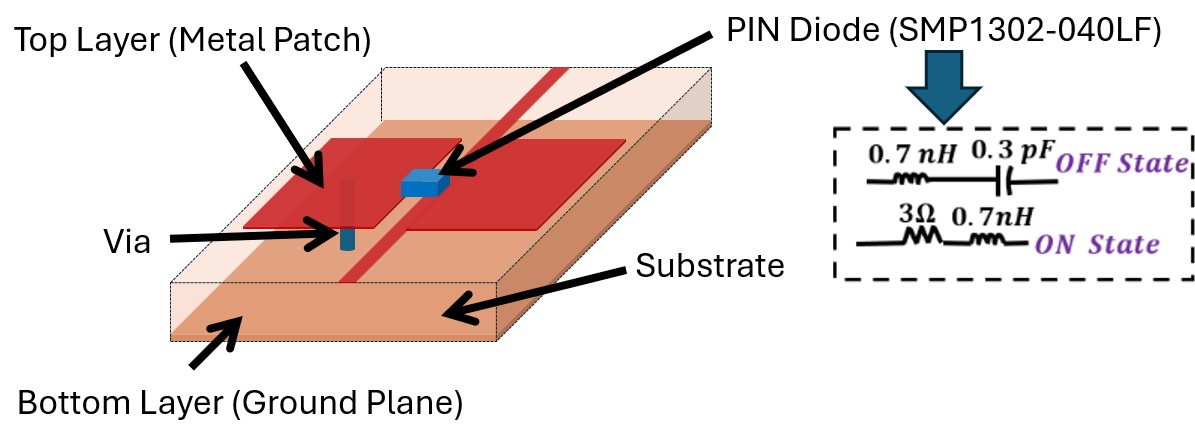}
         \caption{Schematic diagram for the unit-cell of $1$-bit digitally coded RIS: Perspective 3D view and Equivalent circuit of PIN diode~\cite{Malleboina_EuCAP_2025}.}
         \label{unitcell}
         \vspace{-0.1cm}
    \end{figure}

Our work employs a one-bit coded RIS operating at a center frequency of $5$~GHz with a bandwidth of $300$~MHz \cite{Malleboina_EuCAP_2025}. The RIS has a single-layer architecture with comprising of $1024$ unit cells arranged in a $32 \times 32$ uniform planar array (UPA). Each RIS unit-cell has a square dimension of $\lambda_0/4$, where $\lambda_0 = 60$~mm is the free-space wavelength at $5$~GHz. To enable reconfiguration between two states, each unit-cell incorporates an SMP1302–040LF PIN diode (see Fig.~\ref{unitcell}). In the ON state, the PIN diode is represented by a series connection of a resistor and inductor with values $R_{\text{ON}} = 3~\Omega$ and $L_s = 0.7$~nH, respectively. In the OFF state, it is modeled as a series combination of a capacitor and inductor with $C = 0.3$~pF and $L_s = 0.7$~nH. When simulated using periodic boundary condition, these distinct electrical responses of PIN diodes yield reflection phases of $-28^\circ$ (ON state) and $150^\circ$ (OFF state) for the unit-cell, corresponding to a phase difference of approximately $180^\circ$ which is suitable for the implementation of one-bit digital coding. 

Depending on the values of illumination angle and desired beam-steering angle, the reflection-phase gradient required at the RIS is first computed analytically \cite{Raju_AWPL_2023}. Accordingly, the one-bit quantized phase-gradient profile is realized with reflection phase levels of $0^\circ$ and $180^\circ$ at suitable locations.  Full-wave electromagnetic simulations using CST Microwave Studio confirm scan-loss-free beam-steering from the RIS over a range of $0^\circ$–$60^\circ$, achieving a half-power beamwidth (HPBW) of $6^\circ$ and sidelobe levels below $-9$~dB. The phase shifts across the RIS cells are controlled by a field-programmable gate array (FPGA) based controller module, and the beam steering functionality is experimentally verified in an anechoic chamber \cite{Malleboina_EuCAP_2025}. Our one-bit coded metasurface based RIS provides wide-angle beam-steering functionality with finer resolution compared to reflectarray based RIS architectures. The beam-steering of the RIS is based on analytically computed reflection phase distribution which simplifies the codebook realization. Furthermore, our RIS utilizes PIN diodes and radial stub in place of Varactor diodes and on-chip inductors, making it significantly cost-effective and scalable to larger dimensions. 

\vspace{-0.25cm}
\subsection{The OpenAirInterface for 5G NR}\label{sec_OAI_description}

\subsubsection{Testbed Details}\label{sec_testbed_architecture}
The $5$G NR testbed is implemented using the OAI platform~\cite{oai_roadmap2019, oai_platform2014} in a monolithic stand-alone (SA) configuration that supports the full $5$G RAN protocol stack and core network. Universal software radio peripherals (USRPs) serve as radio units (RUs) to enable over-the-air (OTA) communication between the gNB and UEs. The OAI $5$G RAN and UE software stacks run on an Intel i$7$ computers configured with $12$~cores at a speed of $4.9$~GHz, memory of $32$~GB RAM, and Ubuntu $22.04$~LTS with low-latency kernels.

\subsubsection{Communication Resource}
The system operates over a $40$~MHz bandwidth centered at $\approx 5$~GHz. With numerology~$1$ (i.e., $30$~kHz subcarrier spacing), each slot lasts for $0.5$~ms and contains $14$ orthogonal frequency-division multiplexing (OFDM) symbols. A time-division duplexing (TDD) with a downlink (DL) - uplink (UL) periodicity of $10$ slots ($5$~ms) is used. Out of every $10$ slots, $6$ are assigned to DL, $3$ to UL, and $1$ is a mixed slot comprising $6$ DL symbols, $4$ guard symbols, and $4$ UL symbols. The system uses $106$ physical resource blocks (PRBs) across the $40$~MHz band, with $\approx 13$ symbols in each slot allocated for scheduling transport blocks.

\subsubsection{Discrete-Rate Adaptation for Signal Transmission}\label{sec_MCS}
OAI employs discrete-rate adaptation to control the downlink data rate according to channel conditions in each time slot. The rate adaptation is determined by the modulation and coding scheme (MCS), which specifies $a)$ the modulation order and $b)$ the code rate of the forward error correction (FEC) code. Following the $3$GPP defined 64~QAM MCS Index Table~$1$ for physical downlink shared channel (PDSCH)~\cite{3gpp_NR_MCS_table}, there are $29$ choices of MCS index, and higher indices correspond to higher-order constellations, and yield better spectral efficiencies (SE). In OAI, the MCS selection depends on the maximum tolerable BLER at the UE, estimated using a function of channel quality indicator (CQI) (e.g., see~\cite{Lagen_Exp_ICC_2020}). In our setup, the MCS index ranges from $3$ to a CQI-dependent maximum, with the CQI updated every $80$~ms. Following the hybrid automatic repeat request (HARQ) mechanism~\cite{Lagen_Exp_ICC_2020}, the MCS index is increased or decreased by $1$ if the measured BLER (defined as the ratio of the total number of retransmissions to the total number of scheduled transmissions) over a $100$~ms window falls below $0.05$ or exceeds $0.15$, respectively, in accordance with 3GPP specifications. \textcolor{black}{If there is a NACK for any scheduled transmission, then the same transport block can be retransmitted at most 3 times before discarding it, by using the same MCS that was used for the initial transmission.} 

\subsubsection{Resource Allocation via Proportional-Fair Scheduler}\label{sec_PF_scheduler}
In OAI, user scheduling is performed by a PF scheduler that allocates time-frequency resources to maximize the system throughput while ensuring fairness among UEs. At time slot $t$, the BS schedules UE-$k^*(t)$ on all PRBs, according to
\vspace{-0.05cm}
\begin{equation}\label{eq_PF_scheduler_rule}
\vspace{-0.1cm}
    k^*(t) = \argmax\nolimits_{k \in \{1,2,\ldots,K\}} \eta_{k,\textrm{PF}}(t) \triangleq {R_k(t)}\Big/{T_k(t)},
\end{equation}
where $\eta_{k,\textrm{PF}}(t)$,  $R_k(t)$, and $T_k(t)$ denote the PF metric, instantaneous SE, and the average SE of UE-$k$ at time~$t$, respectively. For e.g., the Shannon's formula for $R_k(t)$ is
\vspace{-0.1cm}
\begin{equation}
\vspace{-0.1cm}
    R_k(t) = \log_2\left(1+\left|h_k(t)\right|^2\left({P}\Big/{\sigma^2}\right)\right),
\end{equation}
where $h_k(t)$ is as per~\eqref{eq_overall_ch_UE_k}, $P$ is the BS transmit power, and $\sigma^2$ is the noise variance. The average SE, $T_k(t)$, is computed as an exponential-weighted moving average (EWMA):
\begin{equation}\label{eq_PF_scheduler_Tk_update}
  T_k(t+1) = \left(1-\alpha\right)T_k(t) + \mathbbm{1}_{\{k=k^*(t)\}} \alpha R_k(t)
 \end{equation}
where $\alpha \in (0,1)$ is the EWMA parameter. A small $\alpha$ prioritizes UEs with higher instantaneous SE to be scheduled (which maximizes the system throughput in an opportunistic manner), while larger $\alpha$ promotes fairness by ensuring all UEs are served within shorter time intervals, albeit with lower SE. 
\vspace{-0.45cm}
\subsection{Core Idea and Objectives of this Paper}
\textcolor{black}{Our core idea is to replace explicit RIS optimization step with an approach where the RIS phase configuration is selected randomly and independently over time with an appropriately chosen switching interval between the states. The inherent PF scheduler in $5$G NR then naturally prioritizes transmitting data to the UE that experiences the highest instantaneous RSRP under the current RIS configuration, in turn delivering near-optimal RIS benefits.
In this context, the objective of this paper is to experimentally evaluate the performance of a randomized RIS combined with opportunistic PF scheduling in a real-time $5$G NR setup (see Sec.~\ref{sec_OAI_description}). Specifically, we address:}
\begin{enumerate}
    \item  How much gain does an RIS provide in the channel to a UE compared to a system without an RIS? 
        \item How do the RSRP, MCS index, BLER, and throughput vary across UEs when the RIS configurations are randomly sampled across time and PF scheduler is used?
    \item How does the throughput achieved with a randomly configured RIS under a PF scheduler compare with that obtained using optimized RIS phase configurations, which is considered in traditional RIS-based works?
\end{enumerate}
We next detail the methodology used for answering the above.

\vspace{-0.1cm}
\section{Evaluation Methodology}

\begin{figure}
\vspace{-0.0cm}
    \centering
\includegraphics[width=\linewidth]
{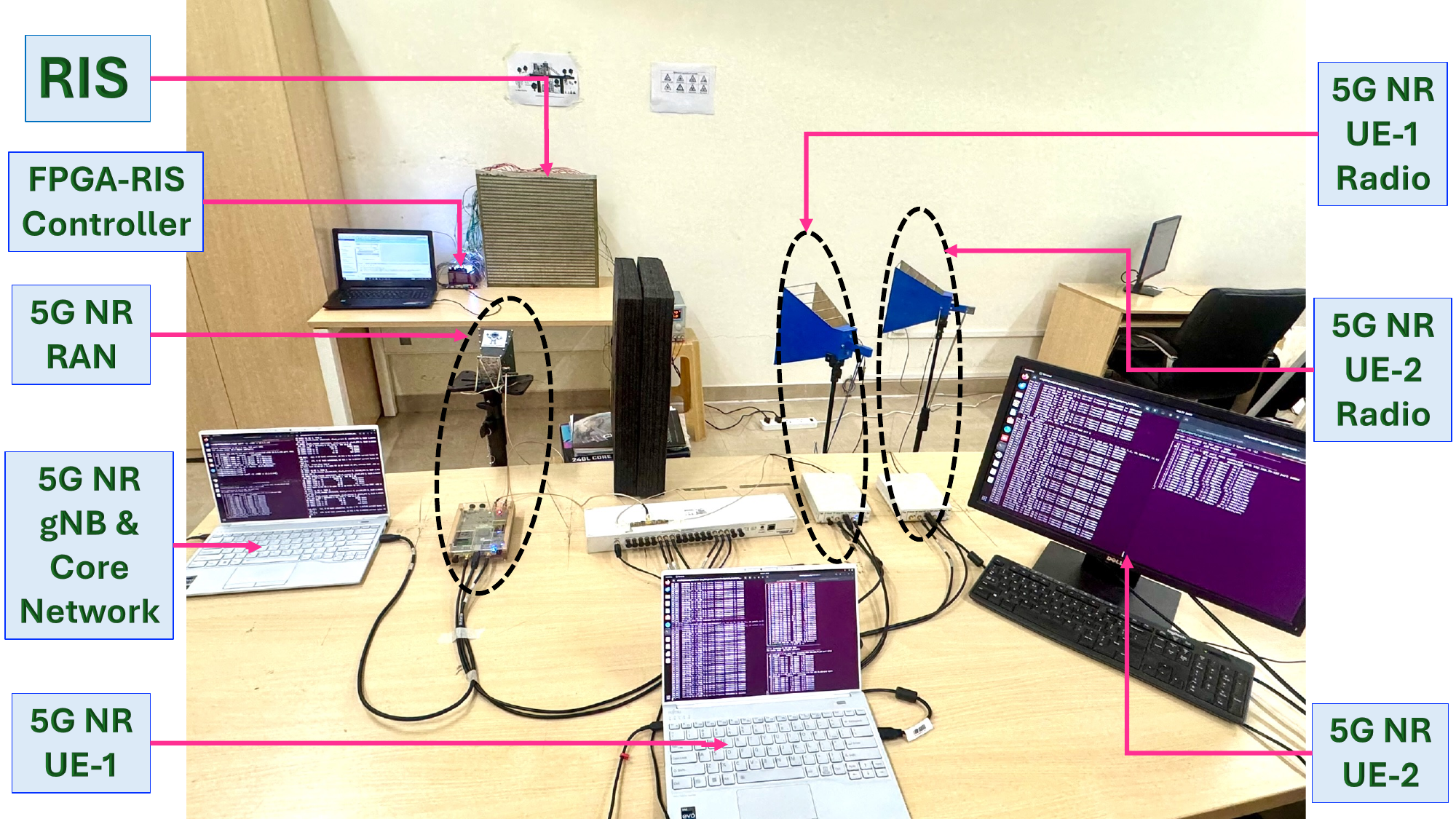}
    \caption{Experimental setup used in this work. The system uses an OAI-based $5$G NR implementation with $1$ gNB and $2$ UEs connected to it via an RIS.} 
    \label{fig:lab_setup}
    \vspace{-0.05cm}
\end{figure}
In this section, we formally describe the proposed approach for randomly configuring the RIS phase and outline the process used to collect the data that support our key findings.
\vspace{-0.15cm}
\subsection{Choice of RIS configurations}\label{Sec_RIS_distribution}
Let $\nu_1,\nu_2,\ldots,\nu_K$ and $\psi_1,\psi_2,\ldots,\psi_K$ denote the azimuth and elevation (cascaded) angles to the $K$ UEs, respectively, at the UPA-based RIS. The RIS phase configuration is chosen from one of $L$ possible states: $\boldsymbol{\phi}_1, \boldsymbol{\phi}_2, \ldots, \boldsymbol{\phi}_L$, where each state $\boldsymbol{\phi}_\ell$ is selected with probability $p_\ell$ such that $p_\ell \geq 0$ and $\sum_{\ell=1}^{L} p_\ell = 1$.
The \emph{sampling distribution} of the RIS states, which is defined by the set of tuples $\left\{\left(\boldsymbol{\phi}_1, p_1\right), \left(\boldsymbol{\phi}_2, p_2\right), \ldots, \left(\boldsymbol{\phi}_L, p_L\right)\right\}$, and the number of states $L$, critically determines the overall system performance. Theoretical insights from~\cite[Theorem~1]{Yashvanth_WCL_2025} provide guidelines for selecting this distribution; it shows that to maximize the system throughput, the RIS phase distribution should match the statistics of the set of channels (which are characterized by the angular parameters ${\nu_1,\nu_2,\ldots,\nu_K,\psi_1,\psi_2,\ldots,\psi_K}$) for which the RIS is intended to deliver optimal gains. 
\vspace{-0.07cm}
\subsection{Experimental Setup and Collection of Data}\label{sec_data_collection}

The experimental setup is shown in Fig.~\ref{fig:lab_setup}. The 5G gNB and UEs are configured as described in Sec.~\ref{sec_testbed_architecture}, using directional horn antennas for RF signal transmission and reception. Further, to focus on the effect of the RIS, the direct BS–UE link is physically blocked. For simplicity, we consider $K=2$ UEs, but our approach extends easily to more users. The UEs are positioned such that $\nu_1 = 30^\circ$, $\nu_2 = 45^\circ$, and $\psi_1 = \psi_2 = 0^\circ$.
From Sec.~\ref{Sec_RIS_distribution}, the sampling distribution that maximizes system throughput requires the RIS to alternate between two distinct phase-shift vector states~\cite[Theorem~1]{Yashvanth_WCL_2025}:
\begin{equation}\label{eq_sweeping_states}
\boldsymbol{\phi}(t) = \begin{cases}
   \boldsymbol{\phi}_1 \triangleq  \mathbf{a}(\nu_1)\otimes\mathbf{a}(\psi_1), \quad \text{with probability } \frac{1}{2},\\
     \boldsymbol{\phi}_2 \triangleq \mathbf{a}(\nu_2)\otimes\mathbf{a}(\psi_2), \quad \text{with probability } \frac{1}{2},
\end{cases}
\end{equation} 
where $\otimes$ denotes the Kronecker product, and $\mathbf{a}(\omega)$ represents the array steering vector of an $N$-element RIS oriented at angle $\omega$ with inter-element spacing $d_{\textrm{RIS}}=\lambda_0/4$, given by
\vspace{-0.1cm}
\begin{equation}
\mathbf{a}(\omega)=\left[1, e^{-j\frac{\pi}{2}\sin(\omega)}, \ldots, e^{-j(N-1)\frac{\pi}{2}\sin(\omega)}\right]^T.
\vspace{-0.1cm}
\end{equation}
The choice of $d_{\textrm{RIS}}=\lambda_0/4$ (c.f. Sec.~\ref{sec_RIS_design}) enables a more compact RIS by reducing the size of a unit-cell and yields lower sidelobe levels compared to $\lambda_0/2$ spacing.
From antenna array signal processing theory, setting the RIS phase-shift vector $\boldsymbol{\phi}$ to $\mathbf{a}(\omega)$ directs the reflected signal energy toward the angle $\omega$, i.e., the RIS \emph{beamforms} toward $\omega$.
\vspace{-0.05cm}
\begin{remark}\label{remark_random_ris}
Although the RIS states in~\eqref{eq_sweeping_states} coincide with those obtained through an optimized design, each state here is chosen independently and with equal probability at every time slot. Consequently, a UE attains optimal beamforming gain only when the randomly selected RIS state aligns with its instantaneous channel. Hence, the RIS is said to be \emph{randomly configured}, indicating that configuring the RIS with any particular phase state is independent of the scheduled UE.
\end{remark}
For real-time collection of data, UEs request downlink data using a transmission control protocol (TCP) over the $5$G NR PDSCH. Measurements are collected under two scenarios:
\begin{itemize}[leftmargin=*]
    \item \textbf{Single-UE scenario}: In this case, only one UE is connected to the gNB at a time. For each UE, experiments are conducted both with and without the RIS. When the RIS is present, it is beamformed towards the respective UE under consideration. For all configurations, we record the time traces of RSRP, BLER, MCS, and \texttt{iperf} throughput.
    \item \textbf{Scheduling scenario}: Both UEs simultaneously connect and request data from the gNB, which employs a PF scheduler to determine the UE for downlink transmission in each time slot. As a practical realization of the RIS states in~\eqref{eq_sweeping_states}, the RIS alternates between the two states with a chosen \emph{switching time} of $T_s = 9$ seconds, independent of the instantaneous channel conditions \textcolor{black}{(for more details on $T_s$, please see Remark~\ref{remark_Ts} in Sec.~\ref{Sec_numerical_final})}. From the perspective of the $5$G gNB and UEs, this appears as a random sequence of phase configurations, as explained in Remark~\ref{remark_random_ris}. We record the RSRP, BLER, MCS, and \texttt{iperf} throughput at the UEs for commonly used values of $\alpha: 0.01, 0.0005,0.00005$.
\end{itemize}

We next present the main experimental findings, and demonstrate that a randomly configured RIS along with an appropriately tuned inherent scheduling mechanisms of a $5$G system, can achieve performance comparable to an optimized RIS and avoids the complexity of real-time phase optimization.
\color{black}

\vspace{-0.05cm}
\section{Main Results: Randomized RIS is Sufficient}
We begin studying the results by quantifying the improvements achieved by deploying an RIS and then analyze the temporal behavior of key performance metrics, including RSRP, BLER, MCS, and throughput, under PF scheduling of UEs.
\vspace{-0.15cm}
\subsection{Quantifying the benefits of the RIS}
Table~\ref{table_with_without_RIS} reports the average RSRP values observed at both UEs under the single-UE scenario, with and without the RIS. 
The RSRP enhances by $\approx 7$-$8$~dB, which corresponds to nearly a $7\times$ increase in received signal power. Since the RSRP directly impacts throughput, both UEs also experience a noticeable increase (by about $20$-$25\%$)  in data rates when the RIS is present, \textcolor{black}{demonstrating the practical gains achievable from an RIS. The observed variation of the data rate with RSRP is in line with any standard 5G NR implementation.}
\addtolength{\topmargin}{0.01in}

\begin{table}[htb!]
\centering
\captionof{table}{Performance with and without RIS.}
\begin{tabular}{|c|cc|cc|}
\hline
\multirow{2}{*}{\begin{tabular}[c]{@{}c@{}} \textbf{With vs.} \\ \textbf{without RIS} \end{tabular}} & \multicolumn{2}{c|}{\textbf{RSRP (in dBm)}}         & \multicolumn{2}{c|}{\textbf{\begin{tabular}[c]{@{}c@{}}Throughput \\ (in Mbps)\end{tabular}}} \\ \cline{2-5} 
& \multicolumn{1}{c|}{UE-$1$}    & UE-$2$    & \multicolumn{1}{c|}{UE-$1$}                              & UE-$2$                             \\ \hline
Without RIS                                                                                     & \multicolumn{1}{c|}{$-112$} & $-110$ & \multicolumn{1}{c|}{$40.70$}                             & $47.40$                            \\ \hline
With RIS                                                                                        & \multicolumn{1}{c|}{$-105$} & $-102$ & \multicolumn{1}{c|}{$50.10$}                             & $58.20$                            \\ \hline
\end{tabular}
\label{table_with_without_RIS}
\vspace{-0.1cm}
\end{table}

\vspace{-0.2cm}
\subsection{Performance of Randomized RIS with PF scheduling}
\begin{figure*}[hbt!]
\vspace{-0.2cm}
\centering
\vspace{-0.2cm}
\begin{subfigure}{.475\linewidth}
\centering
  \includegraphics[width=9cm,height=5.3cm]{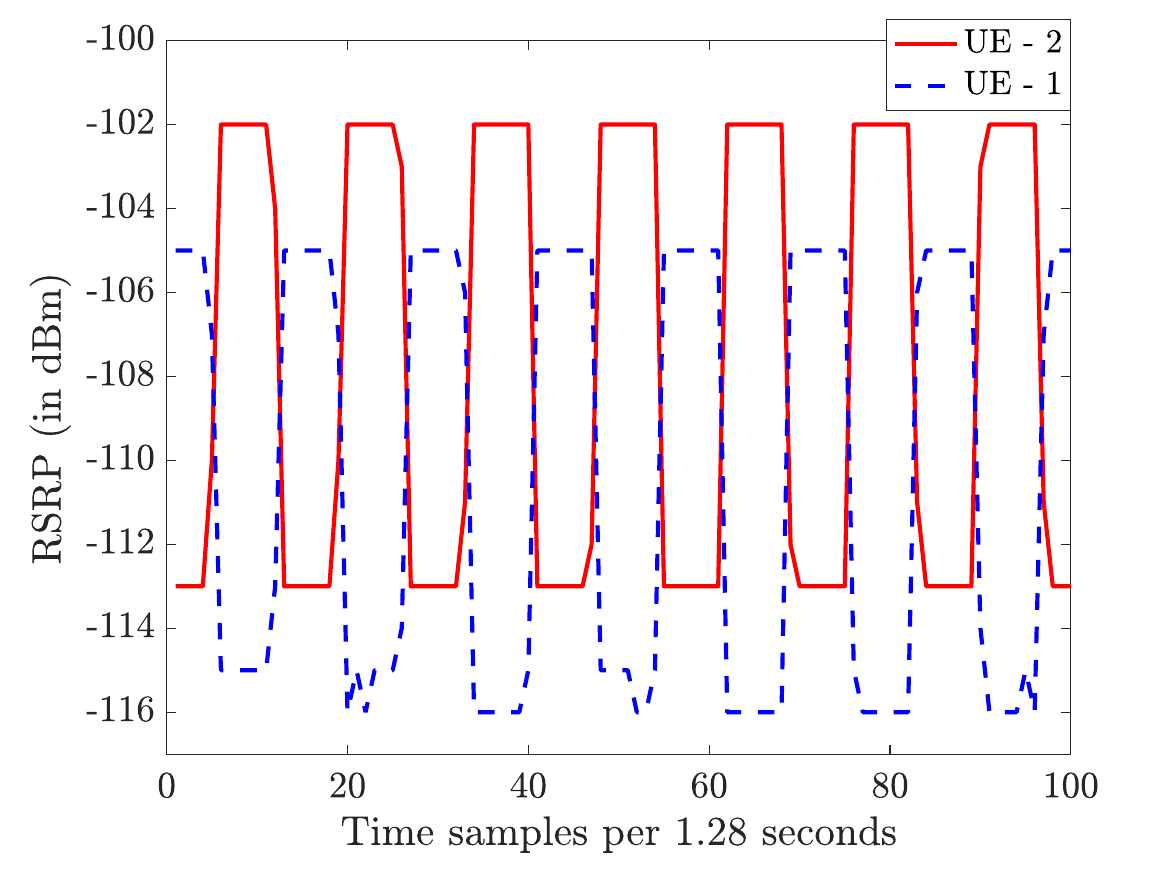}
      \vspace{-0.5cm}
  \caption{RSRP at two UEs.}
  \label{fig_RSRP}
\end{subfigure}
\hspace{0.2cm}
\begin{subfigure}{.475\linewidth}
\centering
  \includegraphics[width=9cm,height=5.3cm]{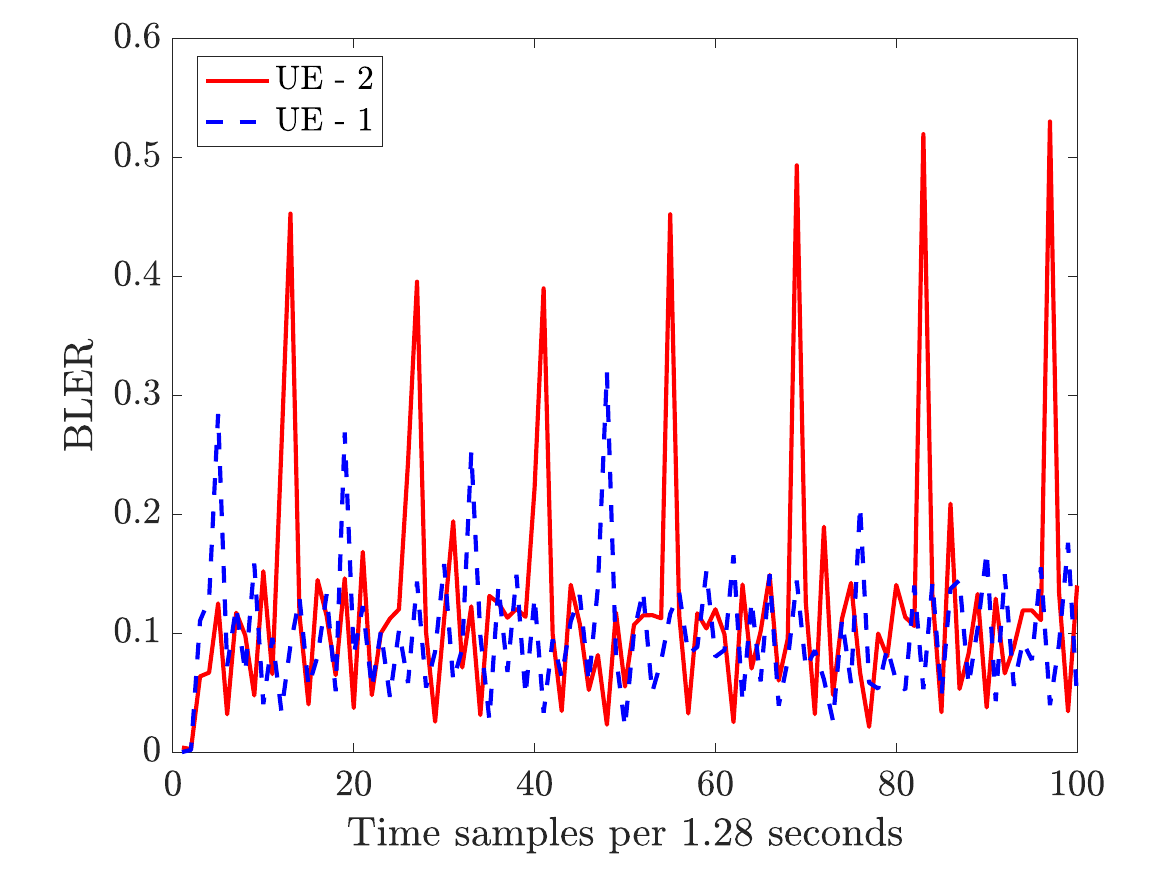}
      \vspace{-0.5cm}
  \caption{BLER at two UEs.}
  \label{fig_BLER}
\end{subfigure}
\begin{subfigure}{.475\linewidth}
\centering
  \includegraphics[width=9cm,height=5.3cm]{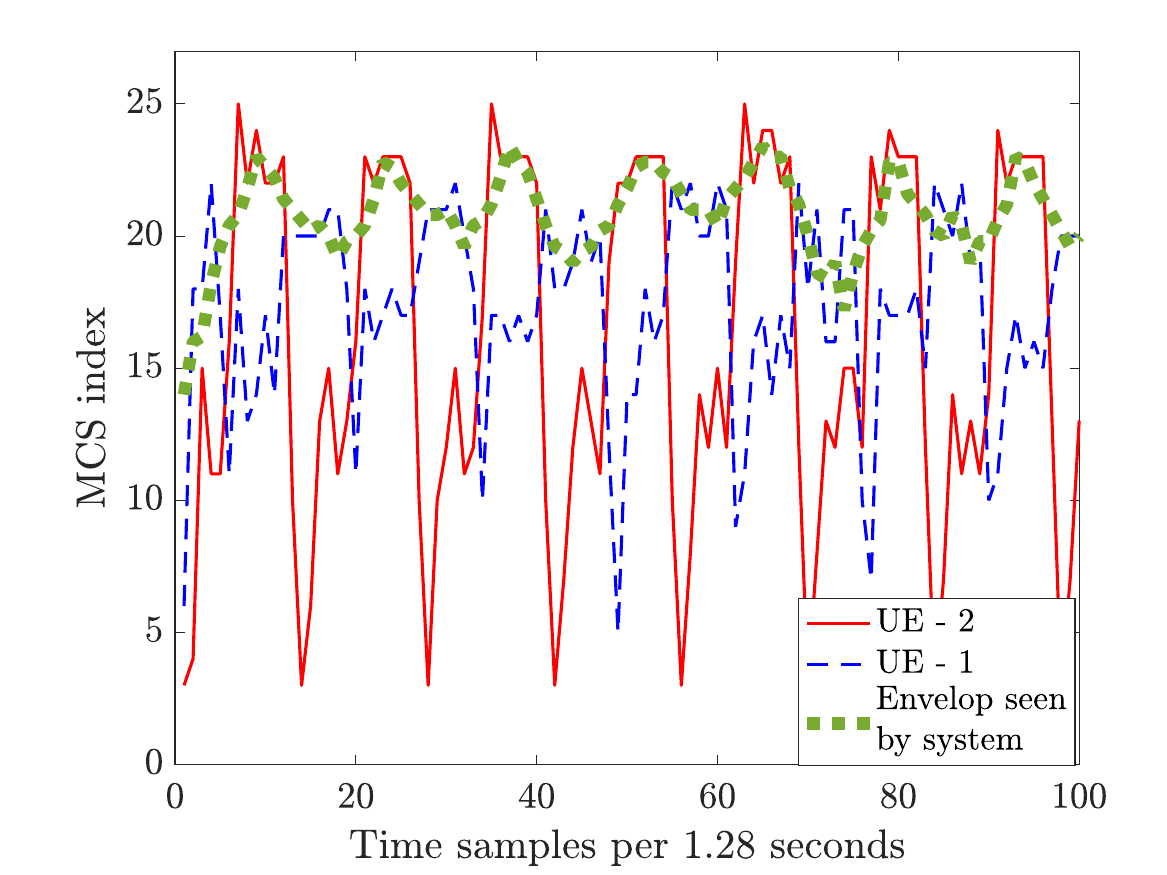}
    \vspace{-0.5cm}
  \caption{MCS indices for two UEs.}
  \label{fig_MCS}
\end{subfigure}
\hspace{0.2cm}
\begin{subfigure}{.475\linewidth}
\centering
  \includegraphics[width=9cm,height=5.3cm]{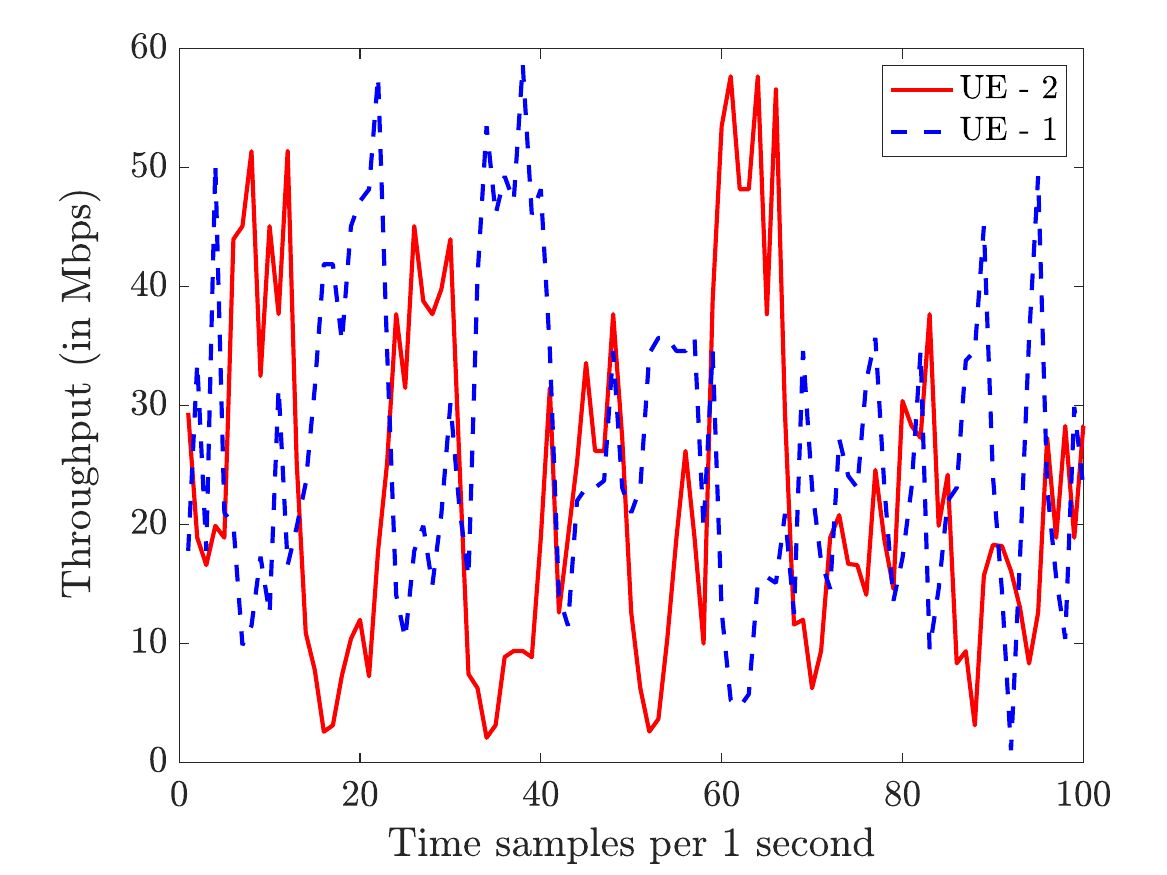}
    \vspace{-0.5cm}
  \caption{Throughput at two UEs.}
  \label{fig_Trpt}
\end{subfigure}
\caption{Variation of performance metrics of a randomized RIS-aided $5$G NR system using a proportional-fair scheduler with $\alpha=0.00005$.}
\label{fig:Time_trace}
\vspace{-0.5cm}
\end{figure*}
We now examine the variation of performance metrics described in Sec.~\ref{sec_data_collection} under the scheduling scenario. Specifically, Fig.~\ref{fig:Time_trace} depicts the temporal evolution of RSRP, BLER, MCS indices, and throughput for both UEs when a PF scheduler operating with $\alpha = 0.00005$ is employed. 

\subsubsection{Variation of RSRP}

Fig.~\ref{fig_RSRP} shows the variation of RSRP at both UEs over time. The RSRP for each UE alternates between high and low levels: $-105$~dBm and $-115$~dBm for UE-$1$, and $-102$~dBm and $-113$~dBm for UE-$2$. This periodic oscillation results from the RIS switching its phase configuration between the beamforming directions of the two UEs, as defined in~\eqref{eq_sweeping_states}. So, when one UE experiences a high RSRP, the other UE has a low RSRP, and vice-versa. 

\subsubsection{Variation of BLER}

In Fig.~\ref{fig_BLER}, we present the variation of BLER experienced by both UEs over time. Recall that a higher BLER indicates that the gNB transmits more information bits than what the channel can reliably support. Accordingly, the MCS is selected such that the BLER at the UEs does not exceed a target value, e.g., $0.1$. Thus, the MCS (and the data rate) dynamically adapts to the channel quality.

From Fig.~\ref{fig_BLER}, it can be observed that, on average, the BLER remains close to the target level of $0.1$, with occasional deviations above and below this value. These deviations can be interpreted as follows. When the RSRP at a UE transitions from a low to a high level, the channel quality improves significantly, thereby allowing the UE to support a higher transmission rate. However, since the gNB initially continues to transmit at the lower rate that was used during the low-RSRP period, the UE is able to decode the received data much more reliably than what is desired for, leading to a temporary drop in the BLER below $0.1$. Conversely, when the RSRP transitions from high to low, the channel degrades while the gNB continues to transmit at the same higher rate that it used during the high-RSRP phase. This mismatch causes the BLER to increase beyond $0.1$ since the transmitted rate now exceeds what the channel can support. Nevertheless, shortly after a transition in RSRP, the gNB adapts its transmission rate to best match the prevailing channel condition, thereby restoring the BLER to its target value of $0.1$. This dynamic rate adaptation ensures that the BLER is maintained around its desired level.

\subsubsection{Variation of MCS Indices}

Fig.~\ref{fig_MCS} illustrates the temporal evolution of the MCS indices at the UEs. As discussed in Sec.~\ref{sec_MCS}, the selection of the MCS index for a given UE depends on the instantaneous channel quality between the gNB and that UE. In an OAI-based $5$G NR, the gNB dynamically adapts the MCS based on whether the observed BLER at the scheduled UE is above or below the target value of $0.1$. 

By jointly examining Figs.~\ref{fig_BLER} and~\ref{fig_MCS}, the following trends can be observed. When the BLER increases beyond $0.1$, which typically occurs when the RSRP decreases from a high to a low level, the MCS index correspondingly drops. This reduction lowers the number of information bits transmitted per symbol, thereby improving decoding reliability and bringing the BLER back toward the target level. Similarly, when the BLER falls below $0.1$ (i.e., when the RSRP transitions from low to high), the gNB raises the MCS index to exploit the improved channel conditions, thus increasing the data rate while maintaining the desired reliability. During time intervals with no change in RSRP, the MCS index is relatively stable at a value that best matches the prevailing channel quality of the scheduled UE.

In Fig.~\ref{fig_MCS}, we also show the envelope of the MCS indices, representing the actual MCS levels achieved under the PF scheduler. Despite the RIS phases being selected randomly, the gNB consistently operates at relatively higher MCS levels. This behavior highlights the effectiveness of the inherent opportunistic scheduling mechanism in $5$G NR, which transmits data to a UE only when it has favorable channel conditions. 

\addtolength{\topmargin}{0.1in}

\subsubsection{Variation of Throughput}
Fig.~\ref{fig_Trpt} presents the time evolution of the throughput achieved by the UEs at the application layer. Since a higher MCS index corresponds to transmitting more information bits per symbol, it directly translates to a higher data rate. Accordingly, consistent with the MCS index variation in Fig.~\ref{fig_MCS}, we observe that whenever the randomly configured RIS happens to direct its beam toward a given UE, that UE experiences a throughput increase. This occurs because the opportunistic scheduler prioritizes data transmission to the UE currently benefiting from the RIS gain, thereby maximizing the system’s instantaneous throughput. This is precisely the well-known \emph{multi-user diversity} effect, wherein the scheduler leverages the channel fluctuations across users to enhance overall performance. Consequently, as will be demonstrated in Fig.~\ref{fig_system_trpt}, this adaptive scheduling mechanism enables the system to achieve a substantially higher aggregate throughput, comparable to that obtained via optimized RIS.
\vspace{-0.1cm}
\subsection{Randomized RIS Can Nearly Procure Full Benefits}\label{Sec_numerical_final}
\begin{figure}
\vspace{-0.2cm}
    \centering
    \includegraphics[width=0.87\linewidth]{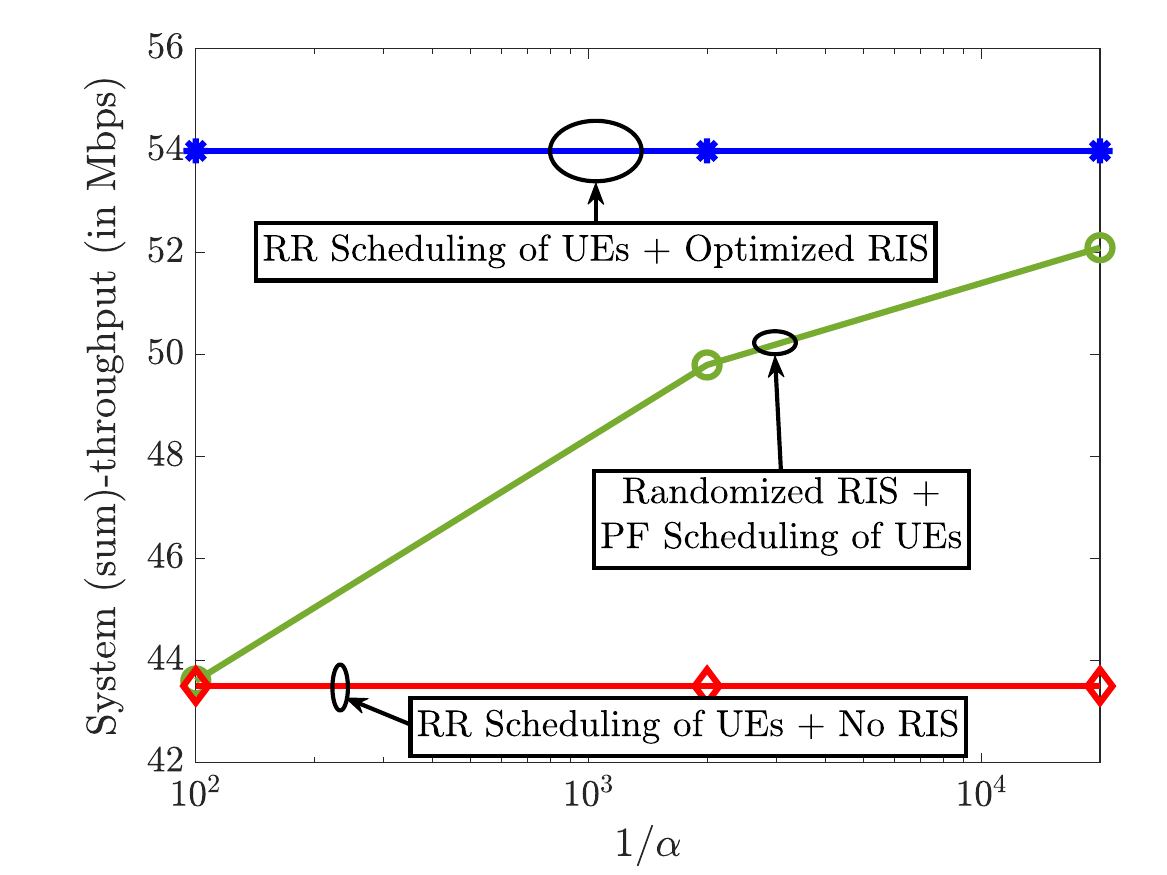}
        \vspace{-0.2cm}
    \caption{System throughput vs. $1/\alpha$.}
    \label{fig_system_trpt}
    \vspace{-0.1cm}
\end{figure}
Finally, we now show that a randomized RIS, when paired with an opportunistic scheduler, can achieve nearly the maximal throughput associated with optimized RIS configuration. Fig.~\ref{fig_system_trpt} depicts the system throughput versus $1/\alpha$, where $\alpha$ is the EWMA parameter in \eqref{eq_PF_scheduler_Tk_update}. The topmost curve depicts the throughput under round-robin (RR) scheduling of UEs with the RIS optimally configured toward the scheduled UE in each slot. This genie-aided curve is obtained by averaging the single-UE throughput values with RIS beamforming to the considered UE (i.e., it is computed from Table~\ref{table_with_without_RIS}); it is independent of $\alpha$ and represents the maximum achievable throughput with an optimized RIS. Practically, achieving this throughput requires tight control of the RIS, configuring it based on the scheduled UE. When the RIS instead switches randomly according to~\eqref{eq_sweeping_states} and the gNB employs PF scheduling, and $\alpha$ is large, the scheduler prioritizes fairness and allocates resources without accounting for the UEs' instantaneous channels, yielding a relatively low throughput. As $\alpha$ decreases, the scheduler increasingly favors selecting the UE with better instantaneous channels, i.e., the UE to which the randomly chosen RIS beam is directed, and the system performance approaches the genie-aided optimized RIS, in line with~\cite{Yashvanth_TSP_2023}.
\emph{Importantly, the figure demonstrates that the throughput achieved with a randomized RIS and a properly tuned PF scheduler at the gNB not only consistently surpasses that of the no-RIS case, but also delivers near optimal RIS gains without explicitly optimizing the RIS.}
\vspace{-0.1cm}
\color{black}
\begin{remark}[Choice of $\alpha$ and $T_s$]\label{remark_Ts}
In general, a judicious selection of the RIS switching interval, $T_s$, is also crucial. Specifically, $T_s$ should be long enough to allow the $5$G NR control signaling and scheduling mechanisms to operate and transmit data to the UE currently benefiting from the near-beamforming RIS gain, yet not so long that the PF scheduler (with a given EWMA parameter $\alpha$) is compelled to serve the other UE to which the RIS is not aligned. Therefore, careful tuning of the $(\alpha, T_s)$ pair is essential to the success of the proposed approach. In this work, $T_s$ was empirically chosen to reveal the benefits of randomized RIS; a formal analysis of their interdependence will be presented in an extended work, where we will theoretically show that $T_s$ scales as $\mathcal{O}(\alpha)$.
\end{remark}
\color{black}
\vspace{-0.15cm}
\begin{figure}
\vspace{-0.3cm}
    \centering
    \includegraphics[width=0.85\linewidth]{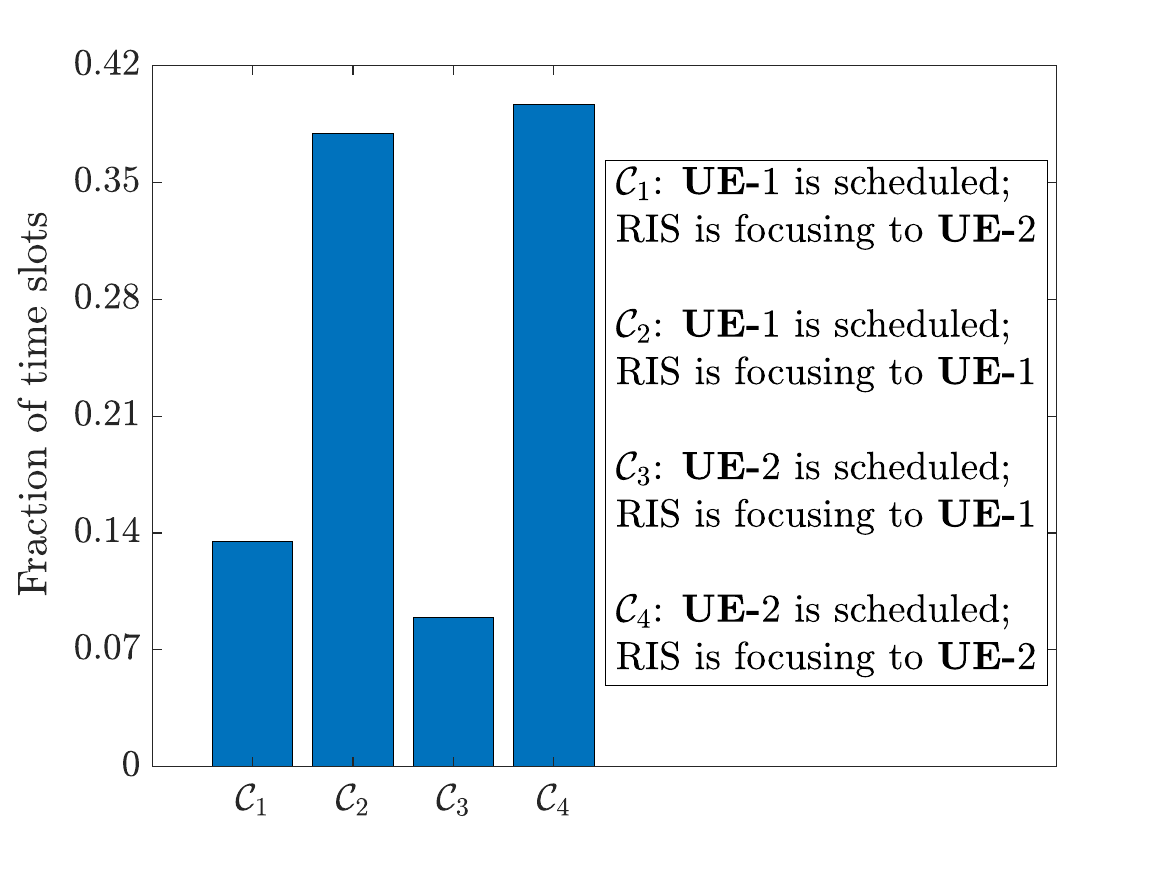}
    \vspace{-0.35cm}
    \caption{Fraction of time slots allotted by PF scheduler with $\alpha=0.00005$.}
    \label{fig_histogram}
    \vspace{-0.25cm}
\end{figure}

\addtolength{\topmargin}{0.01in}

To further substantiate the above findings, Fig.~\ref{fig_histogram} presents a histogram showing the fraction of time slots in which each UE is scheduled during its respective favorable (good) and unfavorable (bad) states, based on a single scheduling instance with $\alpha=0.00005$ observed over a long time trace of channel realizations. Specifically, the figure shows the proportion of time slots in which a PF scheduler selects a UE for transmission when the RIS is aligned with that UE’s channel versus when it is not. We see that, despite the randomized nature of the RIS configurations, the scheduler predominantly transmits data to a UE when the RIS is in the beamforming configuration favoring that UE. 
Occasional instances occur where a UE is scheduled even when the RIS is not aligned toward it, for example, when packets need to be retransmitted (in 5G NR, retransmissions are prioritized above PF scheduling decisions.) 
Moreover, the PF scheduler 
also ensures that the UEs are scheduled almost equally over time, each receiving about $40\%$ of the total time slots during which the RIS focuses its energy towards them, thereby maintaining fairness across UEs. 

\vspace{-0.3cm}
\section{Conclusions}
\vspace{-0.05cm}
We experimentally demonstrated the performance benefits of integrating an RIS with a real-time $5$G system. Primarily, our results show that even a randomly configured RIS can achieve near-optimal gains. By leveraging the inherent opportunistic scheduling of $5$G NR, which allocates resources to UEs with favorable channel conditions, the system attains most of the gains without explicitly optimizing the RIS, thereby enabling low-complexity RIS integration. 
Comprehensive analysis of RSRP, BLER, MCS, and throughput metrics further confirms the efficacy of randomized RIS configurations. While the experiments are conducted in an indoor single-antenna environment with two UEs, our approach directly extends to real-world multiple-antenna deployments with a larger number of UEs; evaluations on large-scale 5G networks along with an analytical study of the impact of the dependence between the RIS switching intervals and the EWMA parameter on system throughput are part of our ongoing work.
\vspace{-0.15cm}
\bibliographystyle{IEEEtran}
\bibliography{IEEEabrv,References}	
\end{document}